\documentclass[twocolumn,showpacs,preprintnumbers,amsmath,amssymb,superscriptaddress]{revtex4}
\usepackage{mathrsfs}

\usepackage{float}
\usepackage{graphicx}
\usepackage{dcolumn}
\usepackage{bm}

\begin{document}


\title{Network inference using asynchronously updated kinetic Ising Model}

\author{Hong-Li Zeng}
\altaffiliation{Email address: \tt {hongli.zeng@tkk.fi}}
\affiliation{Department of Applied Physics, Aalto University,
FIN-00076 Aalto, Finland}

\author{Erik Aurell}
\affiliation{Linnaeus Centre, KTH-Royal Institute of Technology,
SE-100 44 Stockholm, Sweden }\affiliation{Department of Information
and Computer science, Aalto University, FIN-00076 Aalto, Finland}

\author{Mikko Alava}
\affiliation{Department of Applied Physics, Aalto University,
FIN-00076 Aalto, Finland}

\author{Hamed Mahmoudi}
\affiliation{Department of Information and Computer science, Aalto
University, FIN-00076 Aalto, Finland}

\date{\today}

\begin{abstract}
Network structures are reconstructed from dynamical data by
respectively naive mean field (nMF) and Thouless-Anderson-Palmer
(TAP) approximations. For TAP approximation, we use two methods to
reconstruct the network: a) iteration method; b) casting the
inference formula to a set of cubic equations and solving it
directly. We investigate inference of the asymmetric
Sherrington-Kirkpatrick (S-K) model using asynchronous update. The
solutions of the sets cubic equation depend of temperature $T$ in
the S-K model, and a critical temperature $T_c$ is found around 2.1.
For $T<T_c$, the solutions of the cubic equation sets are composed
of 1 real root and two conjugate complex roots while for $T>T_c$
 there are three real roots. The iteration method is convergent only if
the cubic equations have three real solutions. The two methods give
same results when the iteration method is convergent. Compared to
nMF, TAP is somewhat better at low temperatures, but approaches the
same performance as temperature increase. Both methods behave better
for longer data length, but for improvement arises, TAP is well
pronounced.


\end{abstract}


 \maketitle
\section{INTRODUCTION}
A present challenge in biological research is how to deal with the
data originating from the high-throughput technologies. Information
can often convincingly be structured in the form of networks
\cite{Rice}. Vertices on a network are entities and the links with
numbers or other descriptions attached to them are the interactions
between the elements, in, e.g., the biological system
\cite{cell_network, molecular_network, biological_network, metabolic
networks}. On different levels of abstraction, information about the
interactions between each pair of elements is hence useful to
understand the biological system. Finding interactions between
entities from the empirical data is an inverse problem called
'network reconstruction' \cite{Rice, Gardner, Cocco, Roudi_pre,
Roudi_plos, Y. Roudi}.

In this work, we use an idealized system to generate 'empirical'
data with computer, and then try to reconstruct the network
structure of the system, using this test data. The system is the
kinetic Ising model, intended as a proxy for simultaneous recordings
from many neurons. In this setting, symmetric couplings between the
entities are not appropriate, since two neurons will typically not
act on each other in a symmetric way \cite{amit1989}. The properties
of asymmetric neural networks have been studied previously
\cite{Bausch1986, Feigelman1987, Parisi1986}, but not much work has
been done in the context of network reconstruction. Here we extend a
presently reported approach using dynamic mean field theory
\cite{CNS2010,Yasser_preparing} from synchronously updated models to
asynchronously updated models. The analysis closely parallels that
of \cite{Yasser_preparing}, with the difference that data is
continuous in time. The similarities and differences between our
results and \cite{Yasser_preparing} are commented upon in
Conclusion.

Multi-neuron firing patterns can be observed with present
technologies up to thousands of neurons ( recordings on retina
systems ). Schneidman et al. \cite{Schneidman} showed that the
interactions between neuron pairs could be reconstructed using only
the observed firing rates and the pair-wise correlations. Recently,
questions have arisen whether the methods used in \cite{Schneidman}
generalize to other data sets, and if the approximations involved
can be improved or not \cite{Cocco, Roudi_pre, Roudi_plos, Y.
Roudi}. There has also been significant development on the more
theoretical side \cite{Cocco,Mezard,Enzo Marinari, Charles Ollion}.

A theoretical model, which can be used to generate the frequencies
of all possible spiking configurations is the well-known Ising model
\cite{Cocco}. For a system of $N$ neurons, it is characterized by up
to $N^2$ parameters: $N$ external fields, $\theta_i$, on each
individual neuron, and $N(N-1)$ 'links', $J_{ij}$, between each pair
of neurons. In the asymmetric Ising model, $J_{ij}$  is not equal to
$J_{ji}$. And for S-K model, the symmetrized and anti-symmetrized
couplings $J_{ij}^s$ and $J_{ij}^{as}$ are identically independent
Gaussian distributed random variables. The model is entitled
'kinetic' because, except for the fully symmetric case, it does not
correspond to an equilibrium statistical mechanics system.

With the observed average firing rates and all pairwise equal-time
correlations in an empirical data set, maximum entropy models can
find a probability distribution which maximizes the entropy of the
data domain. This condition implies that the samples are drawn
independently from the same distribution. The state of maximum
entropy given is an equilibrium state which has a probability
distribution of Ising form \cite{Shannon}. The quantities $J_{ij}$
and $h_i$ are then Lagrange multipliers to satisfy the constrains
that the ensemble expectation values agree with sample averages in
the data set. If the data is however generated by a dynamics, then
samplings drawn close in time are typically dependent. This is the
extra information which will be used here through the kinetic
inverse Ising reconstruction scheme. For the equilibrium version of
the inverse Ising problem, Yasser Roudi and collaborators review and
investigate several approximation methods \cite{Y. Roudi,
Roudi_plos, Charles Ollion} with the maximum entropy method,
arriving at the general conclusion that all of them are unreliable
in a dynamic setting, if the systems are sufficiently large, and in
most ranges of parameters. Better inference methods on dynamic data
are called for.

A standard approach to sample the equilibrium Ising model is Glauber
dynamics,
which we will describe below. It is however not restricted to
symmetric Ising model, but also well-defined for models with
asymmetric couplings. It is plausible that such a more general
frame-work can describe the underlying system not close to
equilibrium, and with asymmetric couplings, better. Here we are
therefore interested in using kinetic Ising model, typically with
asymmetric couplings, to reconstruct a neural network dynamically.

There are several reasons to consider asynchronous update models
(Glauber dynamics) instead of synchronous update. The first is that
asynchronous updates converge to a stationary state which for
symmetric models in the Boltzmann-Gibbs equilibrium measure, while
neither is necessarily true for synchronous updates. A second reason
is that most plausible applications are naturally asynchronous. For
instance, the expression of gene is not a synchronous process, the
transcription of DNA and the transport of enzymes may take from
milliseconds up to a few seconds. Another example is the refractory
period for neuron in which the neuron cannot respond to input signal
as it is still processing or recovering from the previous input
signal. The period generally lasts for one millisecond
\cite{example_asynchronous}. Besides,
\cite{asynchronous,Klemm_asyn_biological_implication} show that the
biological networks do not have a completely synchronous update. For
these reasons, we have focused on the asynchronous update Glauber
dynamics. For a discussion of synchronous update we refer to
\cite{CNS2010,Yasser_preparing}.

The paper is organized as follows: we describe the asymmetric S-K
model and Glauber dynamics in Sec. II; the inference formula with
nMF and TAP approximation for asynchronous case is derived in Sec.
III; the performances of the inference formula are given in Sec. IV.
Finally, we summarize the work in Sec. V.

\section{Asymmetric S-K model and Glauber dynamics}
The S-K model is a system of $N$ spins, which models $N$ neurons
with binary states ($s_i=1$ for firing state, otherwise $s_i=-1$ ).
It is a fully connected model, i.e., all neurons in the system have
interactions with each other. The interactions $J_{ij}$ between each
pair of neurons have the following form:
\begin{equation}
 J_{ij} = J_{ij}^s + kJ_{ij}^{as},~~~~~~~k\geq0.
\end{equation}
where, $k$ measures the asymmetric degree of these interactions,
$J_{ij}^s$ and $J_{ij}^{as}$ are symmetric $J_{ij}^s = J_{ji}^s$ and
asymmetric matrices $J_{ij}^{as} = -J_{ji}^{as}$, respectively. They
consists both of identically and independently Gaussian distributed
random variables with means 0 and variances:
\begin{equation}
<{J_{ij}^s}^2> = <{J_{ij}^{as}}^2> =\dfrac{J^2}{N}\frac{1}{1+k^2}.
\end{equation}
The self-connections are avoided, i.e., the on-diagonal elements of
$J_{ij}^s$ and $J_{ij}^{as}$ equals 0.

We now define the kinetic Ising model with asynchronous updates. Let
the joint probability distribution of spin states in system at time
$t$ as $p(s_1,...,s_N;t)$, and let the master equation of our model
be written as
\begin{eqnarray}
\frac{d}{dt}p(s_1,...,s_N;t)~~~~~~~~~~~~~~~~~~~~~~~~~~~~~~~~~~~~~~~~~~~~~~~~\nonumber\\
=\sum_i\omega_i(-s_i)p(s_1,...,-s_i,...,s_N;t)
-\sum_i\omega_i(s_i)p(\textbf{s};t).
\end{eqnarray}
where $\omega_i(s_i)$ is the flipping rate, i.e., the probability
for the state of $i$th neuron changes from $s_i$ to $-s_i$ per unit
time. The flipping rates are given by Glauber dynamics as follows:
\begin{equation}
 \omega_i(s_i)=\dfrac{1}{1+\exp[2\beta s_i(\theta_i+\sum_jJ_{ij}s_j)]}.
\end{equation}
where, $\beta$ is the inverse of temperature $T$. For convenience,
define $H_i=\sum_jJ_{ij}s_j+\theta_i$ as the effective field on
neuron $i$ , where $\theta_i$ is the external field of spin $i$. If
the couplings are symmetric (i.e., $J_{ij}^{as}=0$), then the steady
state of the dynamics given by (3) and (4) is $p(s_1,...,s_N)\propto
exp(\beta{\sum_i s_i\theta_i+\sum_{ij}s_is_jJ_{ij}})$. If the
couplings are not symmetric, then (3) and (4) still have a steady
state (under general condition), but this state does not have a
simple description.

With state for each neuron $s_i$, we can naturally define the time
dependent means and correlations as follows:
\begin{eqnarray}
m_i = \langle s_i(t)\rangle.~~~~~~~~~~~~~~~~~~~~~~~~~~~\nonumber\\
C_{ij}(t-t_0) = \langle s_i(t)s_j(t_0)\rangle - m_im_j.~~~~~~~
\end{eqnarray}

From equation (3) and (4), we get the equation of motion for means
and correlations as
\begin{eqnarray}
\dfrac{dm_i}{dt}= m_i + \langle \text{tanh}[\beta
s_iH_i(t)]\rangle.~~~~~~~~~~~~~~~~~~~~~~\nonumber\\
\dfrac{d}{dt}\langle s_i(t)s_j(t_0)\rangle=-\langle
s_i(t)s_j(t_0)\rangle + \langle \text{tanh}[\beta
H_i(t)s_j(t_0)]\rangle.
\end{eqnarray}

For the second equation of eq. (6), the term in the left hand side
and the first term in the right hand side can be solved based on the
empirical data produced by the Glauber dynamics. However, the
calculation of the average value for $\tanh[\beta H_i(t)s_j(t_0)]$
involves all kinds of higher-order correlations and is therefor not
easily expressed only in terms of means and pair-wise correlations.
In order to solve the second equation  in (6), perturbatively
approximations for the second term of the right hand side are
obviously needed. Here, we use the nMF and TAP approximations
respectively to deal with this \textbf{tanh} function.

\section{nMF approximation and TAP approximation}
The simplest method to find out the parameters of the Ising model
from empirical data is the mean-field theory:
\begin{equation}
 m_i=\text{tanh} \beta ( \theta_i+\sum_j J_{ij}m_j)
\end{equation}
Following recent practice, and to distinguish this first level of
approximation from others, we will refer to it as naive mean- field
(nMF). Let $b_i=\theta_i+\sum_j J_{ij}m_j$ and rewrite $H_i$ as
\begin{equation}
H_i=b_i+\sum_j J_{ij}(s_j-m_j) \equiv \sum_j J_{ij}\delta s_j +b_i.
\end{equation}
Expanding the $\textbf{tanh}$ function with respect to $\beta b_i$
in equation (6)
\begin{eqnarray}
\dfrac{d}{dt}\langle s_i(t)s_j(t_0)\rangle+ \langle
s_i(t)s_j(t_0)\rangle~~~~~~~~~~~~~~~~~~~~~~\nonumber\\
= m_im_j + \beta(1-m_i^2)\left( \sum_k J_{ik}\langle\delta
s_k(t)\delta s_j(t_0)\rangle\right).
\end{eqnarray}
and denoting the time difference $t-t_0$ as $\tau$, we have
\begin{equation}
\frac{d}{d\tau}C_{ij}(\tau) + C_{ij}(\tau) = \beta (1-m_i^2)\sum_k
J_{ik}C_{kj}(\tau).
\end{equation}
In the limit $ \tau \rightarrow 0$, we obtain the equation which we
need to infer the network couplings:
\begin{equation}
J=TA^{-1} D C^{-1}.
\end{equation}
where $D= \dot C +C$ and $A_{ij}=\delta_{ij}(1-m_i^2)$.

Equation (11) is a linear matrix equation with respect to $J_{ij}$.
We can solve it directly.

Next, we turn to derive the inference formula with TAP
approximation. If the Onsager term, i.e.,the effect of the mean
value of neuron $i$ on itself via its influence on another neuron
$j$, is taken into account, the TAP equation is \cite{TAP}
\begin{equation}
m_i=\tanh (\beta b_i-m_i\beta^2\sum_{k \neq i}J_{ik}^2(1-m_k^2)).
\end{equation}
With
\begin{equation}
T_i=b_i \pm m_i\beta^2\sum_{k \neq i}J_{ik}^2(1-m_k^2) + \sum_j
J_{ik}\delta s_k .
\end{equation}
and eq. (12), we expand the \textbf{tanh} function in eq. (6) with
respect to
$$\beta b_i-m_i\beta^2\sum_{k \neq i}J_{ik}^2(1-m_k^2)$$ to the third
order and keep the terms only up to the third of $\textbf{J}$. Then
the corresponding TAP inference formula for $J_{ij}$ is obtained,
which is formally the same as in the nMF approximation.
\begin{equation}
J=TA^{-1} D C^{-1}.
\end{equation}
However, matrix $\textbf{A}$ in TAP formula is different
\begin{equation}
A_{ij}=\delta_{ij}(1-m_i^2)\left[1-\beta^2(1-m_i^2)\sum_j
J_{ij}^2(1-m_j^2)\right].
\end{equation}
Eq. (14) is a function of the couplings $\textbf{J}$, and therefor
it is a nonlinear equation for matrix $\textbf{J}$.

We try to solve eq. (14) for $\textbf{J}$ though two approaches. One
way is to solve it iteratively. We start from reasonable initial
values $J_{ij}^0$ and insert them in the right hand side of the
formula. The resulting $J_{ij}^1$ is the solution after one
iteration. This can be again replaced in the right hand side to get
the second iteration results and etcetera ...
\begin{equation}
J^{t+1}=TA(J^t)^{-1}DC^{-1}
\end{equation}

An alternative way is solving it directly, as done for the
synchronous update model in \cite{Yasser_preparing}, casting the
inference formula to a set of cubic equations. For eq. (15), we
denote
\begin{equation}
F_i=\beta^2(1-m_i^2)\sum_{j}J_{ij}^2(1-m_j^2)
\end{equation} and plug it into eq. (14), and then obtain the following equation
for $J_{ij}$:
\begin{eqnarray}
J_{ij}^{\text{TAP}}= \frac{T*V_{ij}}{(1-m_i^2)(1-F_i)}
\end{eqnarray}
where $V_{ij}=[DC^{-1}]_{ij}$. Inserting eq. (18) into eq. (17), we
obtain the cubic equation for $F_{i}$ as:
\begin{equation}
F_i(1-F_i)^2-\frac{\sum_{j}V_{ij}^2(1-m_j^2)}{1-m_i^2}=0.
\end{equation}
With the obtained physical solution for $F_i$, we get the
reconstructed couplings $J^\texttt{TAP}$ as
\begin{equation}
 J_{ij}^{\text{TAP}}=\frac{J_{ij}^{\text{nMF}}}{1-F_i}.
\end{equation}

It is worth mentioning that for the cubic equation (18), we have
three solutions with possible imaginary parts. Here we study the
real roots of the cubic equation and ignore those solutions with
imaginary parts. When three solutions are all real ones, we take the
smallest one.

We introduce $\Delta$ to measure the difference between the
reconstructed network structure and the original true ones, i.e.,
$\Delta$ is the reconstruction error
$$\Delta=\sqrt{\frac{\sum_{i\neq j}(J_{ij}^{re}-J_{ij}^t)^2}{\sum(J_{ij}^t)^2}}.$$
where $J_{ij}^t$ represents the true network couplings and
$J_{ij}^{re}$ for the reconstructed ones.

\section{The performances of nMF and TAP approximation}

As the starting point, we take a look at the number of solutions
given by nMF and TAP approximation. The nMF gives unique solution
while the iteration method of TAP starting from nMF provides 0
solution when the iteration is divergent and 1 solution for
convergence. However, the cubic-equation method of TAP approximation
always contains at least one solution. Denote the constant term of
eq. (19) as $x$,
\begin{equation}
x=-\frac{\sum_j V_{ij}^2(1-m_j^2)}{(1-m_i^2)}
\end{equation}
$x$ is temperature dependent and negative as $0<m_i^2<1$. The cubic
equation (19) has 3 real roots when $-\frac{4}{27}<x<0$. We only
consider the smallest one and indeed it provides the most accurate
$J_{ij}$'s (data are not shown). With $x<-\frac{4}{27}$, eq. (19)
has only one real root and other two complex solutions with
imaginary part which are discarded as they have no physical meaning.
In Fig. 1, we give the fraction of cubic-equation set (19) (as
$i=1,2,...,N$, where N is the system size) which contains three real
solutions. When the set of cubic-equation at given $T$ contains N
real and $2*N$ complex solutions, we say the fraction of three real
roots equals 0 at this temperature point. As shown in Fig. 1, a
transition seems to occur around $T_c=2.1$. For large system size
and $T<2.1$, the solutions for eq. (19) has only one real root while
for $T > 2.1$ 3 real ones. We plot this figure for data length
$L=N*10^6$, so smaller $N$ means shorter data length, that explains
why the curve of $N=20$ is not quite smooth.
\begin{figure}[H]
\centering
\includegraphics[width=0.45\textwidth]{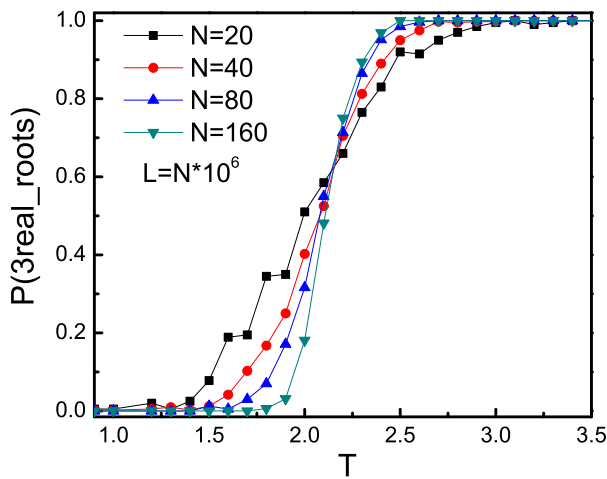}
\caption{(Color online) The fraction of 3 real roots for the cubic
equation set eq. (19). A transition seems to occur around $T_c =
2.1$. Here, we find larger N, the transition curve sharper. The
parameter values: $\theta=0.5$, $k=1$, $L=20*10^6$.} \label{fig4}
\end{figure}

For the simulation of the iteration method of TAP approximation, we
take the reconstructed $J_{ij}^{MF}$ by nMF as the initial input
$J_{ij}^0$, and follow eq. (16) to get $J_{ij}^1$, $J_{ij}^2$...
iteratively. If the average value of $\delta(t)=\overline{\mid
J_{ij}^t-J_{ij}^{t-1}\mid}$ less than the threshold value $10^{-5}$,
then, we consider the iteration is convergent and stop iteration. An
interesting phenomena of the iteration method is it is divergent
when the solutions of cubic-equation set contain complex roots while
convergent when they contain only real roots. Here, we mention three
possible causes for the divergence. One originates from the frozen
states of spin-glass where $m_i^2=1$ and neither nMF nor TAP can
work. A second possible cause: there exists a single fixed point of
the solution but the initial $J_{ij}$'s are drawn as $J_{ij}^{MF}$,
which may a little bit far away from the true solutions for
$J_{ij}$'s at low $T$, and the iteration can not reach to the fixed
point. The last possible cause may come from the fixed point which
is unstable. Here, the given results are for $\theta=0$ and $k=1$,
there is no frozen states for the given temperatures. Then, the
divergence may arise by the second or third possible reason.

\begin{figure}[H]
\centering
\includegraphics[width=0.45\textwidth]{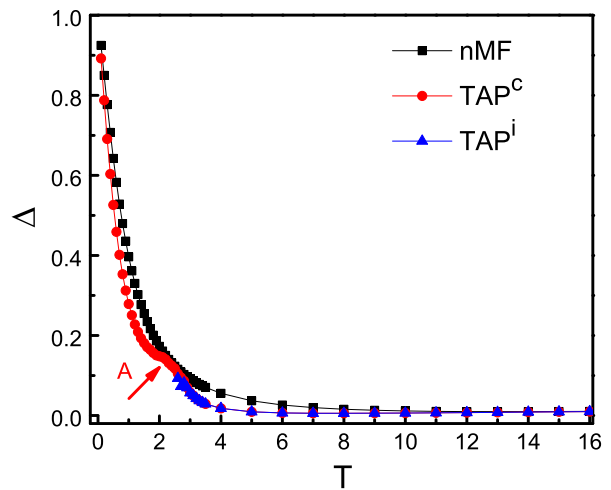}
\caption{(Color online) The reconstruction error $\Delta$ with
temperature $T$ for both nMF and TAP approximation. The other
parameter values: $N=20$, temperature $L=20*10^{10}$, external field
$\theta=0$, asymmetric degree $k=1$. Notations: black square for nMF
red circle for cubic equation method for TAP, blue triangle for
iteration method for TAP. Each data point is averaged on 10
realizations.} \label{fig2}
\end{figure}

We next turn to investigate the influence of $T$ on the
reconstruction errors $\Delta$ in the case of zero external field
($\theta=0$) aS-K model ($k=1$). We plot $\Delta$ with $T$ for nMF
and TAP in Fig. 2. For TAP approximation, when iteration method is
convergent, it produces the same results (blue triangle) as the
cubic-equation method (red circle). Both approximations work better
with temperature $T$ increasing but approach to the same behavior
when $T$ goes higher. It is because for eq. (15), the Onsager term
will approach 0 if $T$ goes high enough, i.e., there will be no
difference between nMF and TAP approximation. As shown in Fig. 2,
TAP always works better than nMF before they approach to the same
results. But there is an noticeable area in which the curve by the
cubic equation method of TAP pointed to with letter 'A' is not as
smooth as that of nMF. The reason is this temperature interval is
located in the critical area where the solutions of the
cubic-equation set eq. (19) are coexistence of two states: some
spins have 3 reals roots and the others have only 1 real root. We
tested also for systems with different size and found that larger
system size give more clear inflexions and closer to the critical
temperature $T_c$, around 2.1. Such results are consistent with the
results shown in Fig. 1.

Fig. 3 illustrates the reconstructions errors for every $J_{ij}$'s
with scatter plots of the inferred $J_{ij}$'s by nMF and TAP
approximation against $J_{ij}^{true}$'s. The left plot is for the
data length $L=N*10^5$ and $L=N*10^7$ for the right one. Here, the
system size $N=20$ and the temperature $T=3.7$ for this plot where
the iteration method of TAP is convergent. The scatter plot shows
that both nMF and TAP perform better for larger $L$. As shown in
both left and right hand side of Fig. 3, the data points for
$J_{ij}^{\texttt{TAP}}$'s inferred by cubic-equation method are
almost covered by that for $J_{ij}^{\texttt{TAP}}$'s inferred by
iteration method, especially for $L=N*10^7$.

\begin{figure}[H]
\centering
\includegraphics[width=0.46\textwidth]{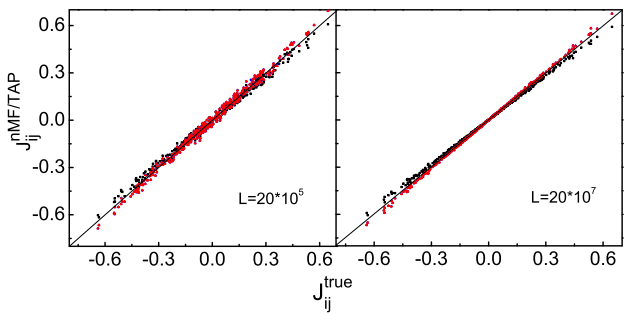}
\caption{(Color online) The scatter plot for the reconstructed
couplings versus the true ones. The parameter values: $N=20$,
temperature $T=3.7$, external field $\theta=0$, asymmetric degree
$k=1$. Notations: black square for inferred couplings using nMF
versus $J_{ij}^{\text{true}}$, blue circle for iteration equation
method of TAP versus $J_{ij}^{\text{true}}$, red triangle for cubic
method of TAP versus $J_{ij}^{\text{true}}$. } \label{fig2}
\end{figure}

From the right plot in Fig. 3, it is difficult to say which
approximation is better as the reconstruction error is quite small
especially with longer data length. Thus, we move next to see how
the data length $L$ works on the reconstruction error $\Delta$ in
the case of zero external field aSK model. With the asynchronously
updating Glauber dynamics, longer data length $L$ ($L=N*L'$, where
$L'$ is the data length in the corresponding synchronous update
case, $N$ is the system size) is needed to obtain comparable results
with that in synchronous case \cite{Yasser_preparing} and say
something about our system. In Fig. 4, $\Delta$ versus $L$ for both
nMF and TAP are plotted for a given temperature $T=8$, where the
iteration method of TAP is convergent. They both reconstruct better
with increasing L, i.e., $\Delta$ decreases as L increases. For
short data length $L<N*10^7$, nMF and TAP produce almost the same
reconstruction error. However, TAP works better than nMF when $L >
N*10^8$. The $\Delta$ for TAP is one order smaller that that for nMF
when $L \geq N*10^9$. Here, again, we find the data points for
cubic-equation method of TAP are covered by those for iteration
method of TAP.
\begin{figure}
\centering
\includegraphics[width=0.45\textwidth]{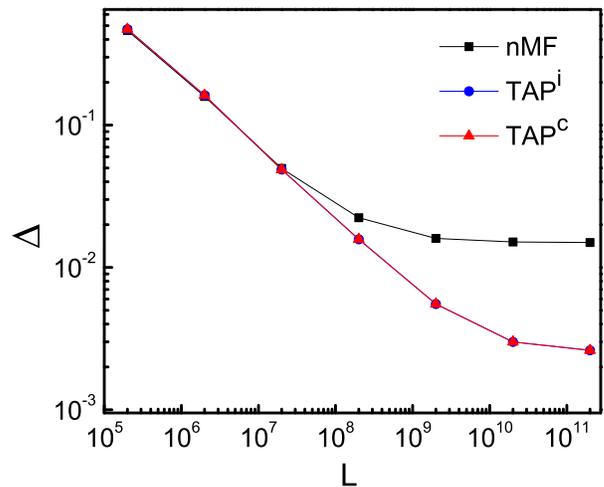}
\caption{(Color online) Reconstruction error $\Delta$ versus the
data length $L$ for nMF and TAP approximation. The other parameter
values: $N=20$, temperature $T=8$, external field $\theta=0$,
asymmetric degree $k=1$. Notations: black square for nMF, blue
circle for iteration equation method of TAP, red triangle for cubic
method of TAP.} \label{fig4}
\end{figure}

The above results are general to different system size $N$. The
performances for nMF and TAP are also compared with non-zero
external field $\theta \neq 0$. We find there exists a frozen state
in the low testing temperature where neither nMF nor TAP can work
there.
\section{Conclusion}
We studied the network inference using asynchronously updated
kinetic Ising model. Two approximations, nMF and TAP, are introduced
to infer the connections and connection strengths in the network. We
have found the transition of the solutions' type for the cubic
equation method of TAP with critical temperature $T_c \approx 2.1$.
We have implemented the TAP approximation as two different schemes,
the cubic scheme, and the iteration scheme. For large system, the
$T_c$ seems to be the starting temperature point for TAP iteration
method to converge.

Comparing our work with \cite{Yasser_preparing} in which the
synchronously updated Glauber dynamics is used, we find two
similarities. The first one is both approximations reconstruct
better with increasing temperature or longer data length. The other
one is TAP works better than nMF especially with long data length at
given temperatures. There are also differences. For instance, the
improvement by TAP approximation in asynchronous case is not as much
as that in synchronous case. Besides, in order to get the comparable
results with synchronous case, the data length for asynchronous case
should be at least $N$ times longer than that for synchronous case.

This work is able to extend to deal with the biological data from
experiments, especially for data produced in continuous time which
correspond to the asynchronous updates. Given the large amount of
data needed to see a difference, we believe that in most application
scenarios, network inference using asynchronously updated kinetic
Ising models should work well enough using naive mean-field (nMF)
reconstruction, and the further step to TAP reconstruction would not
be needed.

\section*{Acknowledgements}
 We are grateful to J. Hertz and Y. Roudi for useful discussions
 about the work and Nordita for hospitality. The work of H.-L. Z.,
 E. A., and H. M. was supported by the Academy of Finland as part of its Finland Distinguished Professor program,
 project 129024/Aurell.

\end{document}